\documentclass[eqsecnum,preprint,preprintnumbers,12pt]{revtex4}
\usepackage{epsfig,psfrag}
\usepackage{hyperref}

\def\nn{\nonumber}

\begin{document}

\title{Calculation of QCD Instanton Determinant with Arbitrary Mass}
\author{Gerald V. Dunne}\email{dunne@phys.uconn.edu}
\affiliation{Department of Physics, University of Connecticut, Storrs, CT 06269, USA
}
\author{Jin Hur} \email{hurjin2@snu.ac.kr}
\author{Choonkyu Lee} \email{cklee@phya.snu.ac.kr}
\affiliation{Department of Physics and Center for Theoretical
Physics\\ Seoul National University, Seoul 151-742, Korea}
\author{Hyunsoo Min} \email{hsmin@dirac.uos.ac.kr}
\affiliation{Department of Physics, University of Seoul, Seoul 130-743, Korea\\ Department of Physics, University of Connecticut, Storrs, CT 06269, USA}

\vskip 2.5 cm

\begin{abstract}
The precise quark mass dependence of the one-loop effective action in an instanton background has recently been computed \cite{idet}. The result interpolates smoothly between the previously known extreme small and large mass limits. The computational method makes use of the fact that the single instanton background has radial symmetry, so that the computation can be reduced to a sum over partial waves of logarithms of radial determinants, each of which can be computed numerically in an efficient manner. The bare sum over partial waves is divergent and must be regulated and renormalized. In this paper we provide more details of this computation, including both the renormalization procedure and the numerical approach. We conclude with  comparisons of our precise numerical results with a simple interpolating function that connects the small and large mass limits, and with the leading order of the derivative expansion.
\end{abstract}
\maketitle

\section{Introduction}
\label{introduction}

The computation of fermion determinants in nontrivial background
fields is an important challenge for both continuum and lattice
quantum field theory. Explicit analytic results are known only for
very simple backgrounds, and are essentially all variations on the
original work of Heisenberg and Euler
\cite{heisenberg,schwinger,brownduff,dunne}. For applications in
quantum chromodynamics (QCD), an important class of background gauge
fields are instanton fields, as these minimize the Euclidean gauge
action within a given topological sector of the gauge field.
Furthermore, instanton physics has many important phenomenological
consequences \cite{thooft,thooft2,shifman,schaefer,diakonov}. Thus,
we are led to consider the fermion determinant, and the associated
one-loop effective action, for quarks of mass $m$ in an instanton
background. Here, no exact results are known for the full mass
dependence, although several terms have been computed analytically
in the small mass \cite{thooft,carlitz,kwon} and large mass
\cite{nsvz,kwon} limits. Recently, in \cite{idet}, the present
authors presented a new computation which is numerical, but
essentially exact, that evaluates the one-loop effective action in a
single instanton background, for any value of the quark mass
(and for arbitrary instanton size parameter). 
The result is fully consistent with the known small and large mass
limits, and interpolates smoothly between these limits. This could
be of interest for the extrapolation of lattice results
\cite{lattice}, obtained at unphysically large quark masses, to
lower physical masses, and for various instanton-based
phenomenology. Our computational method is simple and efficient, and
can be adapted to many other determinant computations in which the
background is sufficiently symmetric so that the problem can be
reduced to a product of one-dimensional radial determinants. While
this is still a very restricted set of background field
configurations, it contains many examples of interest, the single
instanton being one of the most obvious. It is well known how to
compute determinants of \emph{ordinary} differential operators
\cite{levit,coleman,dreyfus,forman,kirsten}; 
but in higher-dimensional problems with \emph{partial} differential
operators, one must confront the renormalization problem since there
are now an infinite number of 1-D determinants to deal with (even
when the partial differential operator has a radial symmetry).

In this paper we present more details of the results of \cite{idet}.
In Section \ref{renormalized} we define the renormalized effective
action in the minimal subtraction scheme, as introduced by 't Hooft
\cite{thooft}, and summarize what is known about the small and large
mass limits. In Section \ref{radial} we review how the single
instanton background reduces the spectral problem to a set of radial
problems, and indicate how to regularize the effective action. This
reduces the computation to two parts, one of which is analytic and
the other is numerical. The analytic part concerns the
renormalization of the effective action, and for this we use a WKB
expansion as is developed in our earlier paper \cite{wkbpaper}. We
stress that this renormalization computation,
which constitutes Section \ref{wkbsection}, 
is analytic and exact, even though we use a WKB expansion, since we
show that only the first two orders of the WKB expansion contribute.
Section \ref{numerical} presents details of the numerical part of
the computation and shows how to combine the numerical part of the
computation with the renormalization part to obtain the finite
renormalized effective action, which is plotted in Figure
\ref{fig5}. In Section \ref{comparison} we present a simple
interpolating function that has been fit to our results, and we also
compare our precise mass dependence with the mass dependence of the
leading derivative expansion approximation, which was computed
previously in \cite{wkbpaper}. The final Section contains some
concluding comments. In Appendix A we give some details (not fully
given in \cite{wkbpaper}) which are needed in the approximate
effective action calculation using the WKB phase-shift method. For
this WKB analysis, the Schwinger proper-time framework
\cite{schwinger} provides a natural way to implement the
renormalization procedure consistently. Appendix B confirms that the same
result is obtained in the regular and singular gauges for the instanton background.

\section{Renormalized Effective Action in a Self-Dual Background}
\label{renormalized}

An instanton background field is self-dual, and self-dual gauge
fields have the remarkable property that the Dirac and Klein-Gordon
operators in such a background are isospectral; that is, they have
identical spectra, apart from an extra degeneracy factor of four in
the spinor case and zero modes present in the spinor case
\cite{thooft,jackiw,brown}. Since the one-loop effective action is
proportional to the logarithm of the determinant of the respective
operator, this has the immediate consequence that it is sufficient
to consider the scalar effective action  to learn also about the
corresponding fermionic effective action, for any mass value m. In
particular, for a quark in a background instanton field, the
renormalized 
one-loop effective action of a Dirac spinor field of mass $m$ (and
isospin $\frac{1}{2}$), $\Gamma^{F}_{\rm ren}(A;m)$, 
can be related to the corresponding scalar effective action, $
\Gamma^{S}_{\rm ren}(A;m)$, 
for a complex scalar of mass $m$  (and isospin $\frac{1}{2}$) by
\cite{thooft,brown, kwon}
\begin{equation}
\Gamma^{F}_{\rm ren}(A;m)= -2\, \Gamma^{S}_{\rm ren}(A;m) 
-\frac{1}{2}\ln\left(\frac{m^2}{\mu^2}\right)\quad , \label{susy}
\end{equation}
where $\mu$ is the renormalization scale. The ln term in (\ref{susy}) corresponds to the existence of a zero
eigenvalue in the spectrum of the Dirac operator for a single
instanton background.

The one-loop effective action must be regularized. We choose
Pauli-Villars regularization adapted to the Schwinger proper-time
formalism, and later we relate this to dimensional regularization, as
in the work of 't Hooft \cite{thooft}. The Pauli-Villars regularized
one-loop scalar effective action is \cite{nsvz,kwon}
\begin{equation}
\Gamma_{\Lambda}^{S}(A;m)  = \ln \left[
{{\rm Det}(-D^2 + m^2) \over {\rm Det} (-\partial^2+m^2)}
{{\rm Det}(-\partial^2 + \Lambda^2) \over {\rm Det} (-D^2+\Lambda^2)}
\label{det}
\right].
\end{equation}
where $D^2 \equiv D_{\mu}D_{\mu}$, with
$D_{\mu}=\partial_{\mu}-iA_{\mu}(x)$. In (\ref{det}), $\Lambda$ is a
heavy regulator mass. We consider an SU(2) single instanton in the
regular gauge \cite{thooft,belavin} :
\begin{eqnarray}
A_{\mu}(x) &\equiv& A_{\mu}^{a}(x)\frac{\tau^{a}}{2}= \frac{\eta_{\mu\nu
a}\tau^{a}x_{\nu}}{r^2+\rho^2},\nonumber\\
F_{\mu\nu}(x) &\equiv& F_{\mu\nu}^{a}(x)\frac{\tau^{a}}{2}
=-\frac{2\rho^2 \eta_{\mu\nu a}\tau^{a}}{(r^2+\rho^2)^2},
\label{instanton}
\end{eqnarray}
where $\eta_{\mu\nu a}$ are the standard 't Hooft symbols \cite{thooft,shifman}.

The regularized effective action (\ref{det}) has  the proper-time representation
\begin{eqnarray}
\Gamma_{\Lambda}^{S}(A;m) &=& - \int_{0}^{\infty}
\frac{ds}{s}(e^{-m^2 s}-e^{-\Lambda^2 s}) \int d^4
x\;\textrm{tr}\langle x|{e^{-s(-{\rm D}^2
)}-e^{-s(-\partial^2 )}}|x\rangle \nonumber \\
&\equiv& - \int_{0}^{\infty} \frac{ds}{s}(e^{-m^2 s}-e^{-\Lambda^2
s}) F(s).
 \label{ptaction}
\end{eqnarray}
The
renormalized effective action, in the minimal subtraction scheme,
is defined as \cite{thooft,kwon}
\begin{eqnarray}
\Gamma^{S}_{\rm ren}(A;m) &=& \lim_{\Lambda\rightarrow\infty}
\left[\Gamma_{\Lambda}^{S}(A;m)-\frac{1}{12} \frac{1}{(4\pi)^2}
\ln \left(\frac{\Lambda^2}{\mu^2}\right) \int d^4
x\;\textrm{tr}(F_{\mu\nu}F_{\mu\nu})\right] \nonumber \\
&\equiv& \lim_{\Lambda\rightarrow\infty}
\left[\Gamma_{\Lambda}^{S}(A;m)-\frac{1}{6}
\ln\left(\frac{\Lambda}{\mu}\right)\right]\quad ,
\label{renaction}
\end{eqnarray}
where we have subtracted the charge renormalization counter-term, and $\mu$ is the renormalization scale.
By dimensional considerations, we may
introduce the modified scalar effective action
$\tilde{\Gamma}^{S}_{\rm ren}(m\rho)$, which is a function of $m\rho$ only, defined by
\begin{equation}
\Gamma^{S}_{\rm ren}(A;m)=\tilde{\Gamma}^{S}_{\rm ren}(m\rho)+ \frac{1}{6}\ln(\mu\rho)\quad ,
\label{modaction}
\end{equation}
and concentrate on studying the $m\rho$ dependence of
$\tilde{\Gamma}^{S}_{\rm ren}(m\rho)$. Then there is
no loss of generality in our setting the instanton scale $\rho=1$ henceforth.

\begin{figure}[tp]
\psfrag{ver}{\Large $\tilde\Gamma_{\rm ren}(m)$}
\psfrag{hor}{\Large $m$}
\includegraphics[scale=1.75]{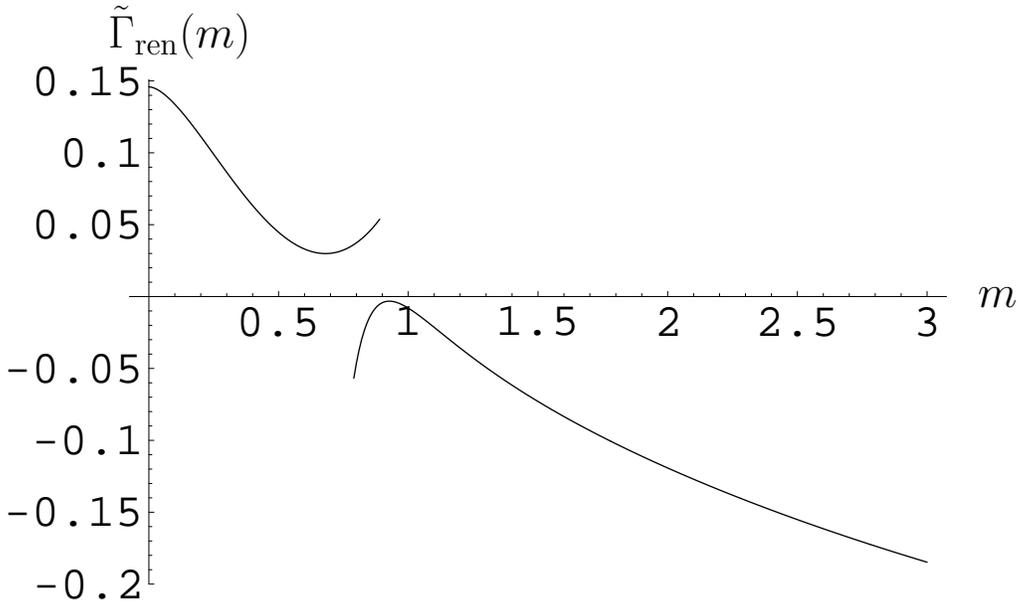}
\caption{Plot of the analytic small and large mass expansions for $\tilde{\Gamma}^{S}_{\rm ren}(m)$, from Equation (\protect{\ref{masslimit}}) Note the gap in the region $0.5\leq m \leq 1$, in which the two expansions do not match up..
\label{fig1}}
\end{figure}

It is known from previous works that in the small mass
\cite{thooft,carlitz,kwon} and large mass \cite{nsvz,kwon} limits,
$\tilde{\Gamma}^{S}_{\rm ren}(m)$ behaves as
\begin{eqnarray}
\tilde{\Gamma}^{S}_{\rm ren}(m)&=&
\begin{cases}
{\alpha\left(\frac{1}{2}\right)+\frac{1}{2}\left(\ln m+\gamma-\ln
2\right)m^2 +\dots \quad , \quad m\rightarrow 0 \cr ~\cr
\displaystyle {
 -\frac{\ln m}{6}-\frac{1}{75 m^2}-\frac{17}{735 m^4}+\frac{232}{2835 m^6}-\frac{7916}{148225 m^8}+\cdots \quad , \quad m\rightarrow \infty}}
\end{cases}
\label{masslimit}
\end{eqnarray}
where
\begin{eqnarray}
\alpha\left(\frac{1}{2}\right)&=& -\frac{5}{72}-2 \zeta^\prime(-1)-\frac{1}{6}\ln 2  \nonumber \\
&\simeq& 0.145873...\quad ,
\label{alphahalf}
\end{eqnarray}
and $\gamma\simeq 0.5772\dots$ is Euler's constant
\cite{abramowitz}. The leading behavior of the small mass limit in
(\ref{masslimit}) was first computed by 't Hooft \cite {thooft}, and
the next corrections were computed in \cite{carlitz} and
\cite{kwon}. This small mass expansion is based on the fact that the
massless propagators in an instanton background are known in
closed-form \cite{bccl}. On the other hand, the large mass expansion
in (\ref{masslimit}) can be computed in several ways. The
$O(\frac{1}{m^2})$ and $O(\frac{1}{m^4})$ terms were computed in
\cite{nsvz}, while the next two terms were computed in \cite{kwon}.
A very direct approach is to use the small-$s$ behavior of $F(s)$,
the proper-time function appearing in (\ref{ptaction}), as given by
the Schwinger-DeWitt expansion. In our case this expansion reads
\cite{kwon}
\begin{equation}
s\rightarrow 0+ \quad :\quad F(s)\sim
-\frac{1}{12}+\frac{1}{75}s+\frac{17}{735}s^2-\frac{116}{2835}s^3+
\frac{3958}{44675}s^4+ \cdots, \label{smalls}
\end{equation}
and using this series with (\ref{ptaction}) immediately leads to the
large $m$ expansion for $\tilde{\Gamma}^{S}_{\rm ren}(m)$ in
(\ref{masslimit}).

Equation (\ref{masslimit}) summarizes what is known analytically about the mass dependence of the renormalized one-loop effective action in an instanton background. This situation is represented in Figure \ref{fig1}, which shows a distinct gap approximately in the region $0.5 \leq m\leq 1$, where the small and large mass expansions do not match up. In this paper we present a technique which computes $\tilde{\Gamma}^{S}_{\rm ren}(m)$ numerically for {\it any} value of the mass $m$. Our results interpolate smoothly between the analytic small and large mass limits depicted in Figure \ref{fig1}.

\section{Radial Formulation}
\label{radial}

Our computational approach makes use of the fact that the single instanton background (\ref{instanton}) has radial symmetry \cite{thooft}. This has the important consequence that the computation of the regularized one-loop effective action (\ref{ptaction}) can be reduced to a sum over partial waves of logarithms of determinants of radial ordinary differential operators. Each such radial determinant can be computed by a simple numerical method, described in Section \ref{numerical}. The physical challenge is to renormalize the (divergent) sum over partial waves.

In the instanton background (\ref{instanton}), with scale $\rho=1$, the Klein-Gordon operator $-D^2$ for isospin $\frac{1}{2}$ particles can be cast in the radial form \cite{thooft}
\begin{equation}
-D^2\to  {\cal H}_{(l,j)} \equiv \left[ - \frac{\partial^2}{\partial
r^2}-\frac{3}{r}\frac{\partial}{\partial
r}+\frac{4l(l+1)}{r^2}+\frac{4(j-l)(j+l+1)}{r^2+1}-\frac{3}{(r^2+1)^2}
\right]\, ,
\label{insth}
\end{equation}
where $l=0,\; \frac{1}{2},\;1,\;\frac{3}{2},\;\cdots\;$, and $j=|\;l
\pm \frac{1}{2}\;|$, and there is a degeneracy factor of
$(2l+1)(2j+1)$ for each partial wave characterized by $(l,
j)$-values. [Note that $l(l+1)$ can be identified with the
eigenvalue of $\vec{L}^2 \equiv L_a L_a$ for $L_a = -\frac{i}{2}
\eta_{\mu\nu a} x_\mu \partial_\nu$, and $j(j+1)$ with the
eigenvalue of $\vec{J}^2 \equiv (L_a +T_a )(L_a + T_a )$ for $T^a =
\frac{\tau^a}{2}$]. In the absence of the instanton background, the
free operator is
\begin{equation}
-\partial^2\to {\cal H}^{\rm free}_{(l)} \equiv \left[ - \frac{\partial^2}{\partial
r^2}-\frac{3}{r}\frac{\partial}{\partial r}+\frac{4l(l+1)}{r^2}
\right] .
\label{freeh}
\end{equation}

This radial decomposition means that we can express the
Pauli-Villars regularized effective action (\ref{det}) also as
\begin{eqnarray}
\Gamma_\Lambda^S(A; m)  &=& \sum_{l=0,\frac{1}{2}, \dots} (2l+1)(2l+2) \left\{ \ln \det \left(\frac{{\mathcal H}_{(l,l+\frac{1}{2})}+m^2}{{\mathcal H}^{{\rm free}}_{(l)}+m^2}\right)+  \ln \det \left(\frac{{\mathcal H}_{(l+\frac{1}{2},l)}+m^2}{{\mathcal H}^{{\rm free}}_{(l+\frac{1}{2})}+m^2}\right)\right.
\nonumber\\
&&\hskip 2cm \left. - \ln \det \left(\frac{{\mathcal H}_{(l,l+\frac{1}{2})}+\Lambda^2}{{\mathcal H}^{{\rm free}}_{(l)}+\Lambda^2}\right)-  \ln \det \left(\frac{{\mathcal H}_{(l+\frac{1}{2},l)}+\Lambda^2}{{\mathcal H}^{{\rm free}}_{(l+\frac{1}{2})}+\Lambda^2}\right)\right\}
\label{pv}
\end{eqnarray}
Here we have combined the radial determinants for
$(l,j=l+\frac{1}{2})$ and
$(l+\frac{1}{2},j=(l+\frac{1}{2})-\frac{1}{2})$, which have the
common degeneracy factor $(2l+1)(2l+2)$, so that the sum over $l$
and $j$ reduces to a single sum over $l$. In our actual analysis,
as explained in detail below, we
need to consider only a truncated sum over $l$ with the expression
(\ref{pv}), and hence possible ambiguities as regards effecting the
infinite sum over $l$ become irrelevant.

In Section \ref{numerical} we present a simple and efficient numerical technique for computing each of the radial determinants appearing in (\ref{pv}). But to extract the renormalized effective action we need to be able to consider the $\Lambda\to\infty$ limit in conjunction with the infinite sum over $l$. This can be achieved as follows. Split the $l$ sum in (\ref{pv}) into two parts as :
\begin{eqnarray}
\Gamma_\Lambda^S(A; m)&=& \sum_{l=0,\frac{1}{2},\dots}^L \Gamma_{\Lambda, (l)}^S(A; m) + \sum_{l=L+\frac{1}{2}}^\infty \Gamma_{\Lambda, (l)}^S(A; m)
\label{actionsplit}
\end{eqnarray}
where $L$ is a large but finite integer. In the first sum, which is finite, the cutoff  $\Lambda$ may be safely removed since for any given finite $l$ \cite{levit,kirsten},
\begin{eqnarray}
&&\lim_{\Lambda\to\infty}  \det \left(\frac{{\mathcal H}_{(l,l+\frac{1}{2})}+\Lambda^2}{{\mathcal H}^{{\rm free}}_{(l)}+\Lambda^2}\right)=1\quad , \nonumber\\
&&\lim_{\Lambda\to\infty} \det \left(\frac{{\mathcal H}_{(l+\frac{1}{2},l)}+\Lambda^2}{{\mathcal H}^{{\rm free}}_{(l+\frac{1}{2})}+\Lambda^2}\right) =1\quad .
\label{remove}
\end{eqnarray}
Thus, the first sum in (\ref{actionsplit}) may be written without the regulator $\Lambda$ as
\begin{eqnarray}
\sum_{l=0,\frac{1}{2},\dots}^L \Gamma_l^S(A; m) & =&\label{lj}\\
&&\hskip -2cm  \sum_{l=0,\frac{1}{2}, \dots}^L (2l+1)(2l+2) \left\{ \ln \det \left(\frac{{\mathcal H}_{(l,l+\frac{1}{2})}+m^2}{{\mathcal H}^{{\rm free}}_{(l)}+m^2}\right)+  \ln \det \left(\frac{{\mathcal H}_{(l+\frac{1}{2},l)}+m^2}{{\mathcal H}^{{\rm free}}_{(l+\frac{1}{2})}+m^2}\right)\right\}\nonumber
\end{eqnarray}
This sum can be computed numerically, and we find [see Section \ref{numerical}] that for any mass $m$ it is quadratically divergent as $L\to\infty$. This divergence is canceled by a divergence of the second sum in (\ref{actionsplit}) in the large $L$ limit, as we show in the next Section.

\section{WKB Analysis and Renormalization}
\label{wkbsection}

In the second sum in (\ref{actionsplit}) we cannot take the large
$L$ and large $\Lambda$ limits blindly, as each leads to a
divergence. However, we show in this Section that the $\Lambda$
divergence is precisely of the counterterm form in the renormalized
action in (\ref{renaction}), and that the large $L$ divergence is
such that it precisely cancels the large $L$ divergences from the
large $L$ limit of the sum in (\ref{lj}). Thus we obtain a finite
renormalized effective action.

The advantage of this technique is that the large $\Lambda$ and
large $L$ divergences of the second sum in (\ref{actionsplit}) can
be computed {\it analytically}, using the WKB approximation for the
corresponding determinants. The WKB approach to radial determinants
was derived in \cite{wkbpaper} up to third order in the WKB
approximation, and the relevant results are reviewed below, and in the Appendix.
It turns
out that in the large $\Lambda$ and large $L$ limits we only need up
to the second order in WKB. The large $L$ limit of the second sum in
(\ref{actionsplit}) can then be analyzed using the Euler-Maclaurin
summation formula \cite{bender}.

It is convenient to express the second piece by the proper-time
representation
\begin{equation}
\sum_{l=L+\frac{1}{2}}^\infty \Gamma_{\Lambda, (l)}^S(A; m)
=\int_0^\infty ds \left[-\frac{1}{s}\left(e^{-m^2 s}-e^{-\Lambda^2
s}\right) F_L(s) \right], \label{second}
\end{equation}
with
\begin{eqnarray}
F_L(s) &=& \int_0^\infty dr \left(\sum_{l=L+\frac{1}{2}}^\infty
f_l(s,r)\right), \label{wkbEM} \\ 
f_l (s,r) &=& (2l+1)(2l+2) [ f_{(l, l+\frac{1}{2})}(s,r) +
f_{(l+\frac{1}{2}, l)}(s,r) ]. \label{wkbfl}
\end{eqnarray}
Here the term $\int_0^\infty dr f_l (s,r)$, obviously related to
$\Gamma^S_{\Lambda,(l)}(A; m)$, can be found using the scattering
phase shifts of the Schr\"{o}dinger problem with the radial
Hamiltonian in (\ref{insth}). Also, in considering the above
infinite sum over partial wave terms, it is now crucial to have the
contribution from partial-wave $(l, j=l+\frac{1}{2})$ and that from
partial wave $(l+\frac{1}{2}, l)$ treated together as a package, as
indicated in (\ref{wkbfl}). For details on this, readers may consult
the Appendix. We here only note that, for large $l$, the WKB
approximation becomes exact and so in the large $L$ limit we can use
the WKB approximation (to an appropriate order) for $F_L (s)$.
According to the WKB expressions derived in the Appendix, it is found
that, for each $l$, $f_l (s,r)$ has a local expansion in terms of
the Langer-modified \cite{langer} potential, $\tilde{V}_{(l,j)}(r)$,  and the corresponding
Langer-modified
free potential, $\tilde{V}_{l}(r)$ :
\begin{eqnarray}
\tilde{V}_{(l,j)}(r) &\equiv&
\frac{4(l+\frac{1}{2})^2}{r^2}+\frac{4(j-l)(j+l+1)}{r^2
+1}-\frac{3}{(r^2 +1)^2}, \label{pot} \\
\tilde{V}_{l}(r) &\equiv& \frac{4(l+\frac{1}{2})^2}{r^2}.
\label{pot0}
\end{eqnarray}
Specifically, the first order WKB result for $f_l (s,r)$ is obtained
by using the form
\begin{eqnarray}
f_{(l,j)}^{(1)}(s,r)=\frac{1}{2\sqrt{\pi s}}\, \exp\left[-s
\tilde{V}_{(l,j)}(r)\right] - \frac{1}{2\sqrt{\pi s}}\,
\exp\left[-s \tilde{V}_{l}(r)\right] \,\, . \label{wkbfirst}
\end{eqnarray}
in the right hand side of (\ref{wkbfl}), and the second order WKB by
using
\begin{eqnarray}
f_{(l,j)}^{(2)}(s,r)&=&\frac{1}{2\sqrt{\pi s}} \left(\frac{s}{4
r^2}-\frac{s^2}{12} \frac{d^2 \tilde{V}_{(l,j)}}{dr^2}\right)
\, \exp\left[-s \tilde{V}_{(l,j)}(r)\right]\nonumber\\
&& -\frac{1}{2\sqrt{\pi s}} \left(\frac{s}{4 r^2}-\frac{s^2}{12}
\frac{d^2 \tilde{V}_{l}}{dr^2}\right) \, \exp\left[-s \tilde{V}_{l}(r)\right]\,\, .
 \label{wkbsecond}
\end{eqnarray}
The corresponding expression for the third order of WKB is also given in
the Appendix, but this result is not needed for our present purposes.

An important observation (which holds true to any order in the WKB
expansion) is that the $l$ dependence in $f_{(l,j)}(s,r)$ has the
form of a polynomial in $l$ multiplied by an exponential in which
$l$ appears quadratically. Thus, if we use the the Euler-Maclaurin
expansion \cite{bender} for (\ref{wkbEM})
\begin{eqnarray}
\sum_{l=L+\frac{1}{2}}^\infty f_l\, = 2 \int_L^\infty dl\, f(l)-\frac{1}{2} f(L)-\frac{1}{24}f^\prime(L)+\dots
\label{euler}
\end{eqnarray}
all terms in this expansion, including the integral term, can be computed analytically.  The integral term yields an error function \cite{abramowitz}
\begin{eqnarray}
\int_L^\infty  dl \exp\left[-a\, l^2 -2 b\, l \right]=\sqrt{\frac{\pi}{4a}}\, e^{b^2/a}\, {\rm Erfc}\left(\frac{a L+b}{\sqrt{a}}\right)\quad ,
\label{efrc}
\end{eqnarray}
where $Re(a)>0$. We here remark that it is important to perform the
$l$-sum prior to considering the $r$-integration.
For more details on the issue of
the integration order, see the Appendix.

To compute the sum in (\ref{second}) we still need to perform the
proper-time integral over $s$ as well as the radial integral over
$r$ appearing in the WKB expression (\ref{wkbEM}). To achieve
this, we adopt the following procedure. First, we trade the
regulator mass $\Lambda$ for a dimensional regularization parameter
$\epsilon$, by demanding that
\begin{eqnarray}
\int_0^\infty ds \left[-\frac{1}{s}\left(e^{-m^2 s}-e^{-\Lambda^2
s}\right) F_L(s) \right] = \int_0^\infty ds \left[-\frac{1}{s}\left(
e^{-m^2 s}\,s^{\epsilon}\right) F_L(s) \right]. \label{dimreg}
\end{eqnarray}
Then, from the facts that $F_L (s) \sim F(s)$ as $s \to 0+$, and $F(s)
= \frac{1}{12} + O(s)$ for small $s$, we see that (\ref{dimreg})
requires
\begin{equation}
-\frac{1}{6} \ln \left( \frac{\Lambda}{m} \right) =
-\frac{1}{12\epsilon} + \frac{1}{12} ( \gamma + 2\ln m) +
O(\epsilon).
\end{equation}
Thus, the correspondence between $\epsilon$ and $\Lambda$ is
\begin{eqnarray}
\epsilon \leftrightarrow \frac{1}{\gamma+2\ln \Lambda}. \label{corr}
\end{eqnarray}

We now proceed to do the proper-time and radial integrals as
follows. First, for $L$ very large, it becomes convenient to rescale
variables as $s\to y/L^2$ and $r\to x\sqrt{y}$. Second, we expand
all terms (except the  $e^{-m^2 s}$ factor) in decreasing powers of
large $L$. Then the $y$ integral can be performed in closed form,
yielding incomplete gamma functions \cite{abramowitz}. These can be
further expanded for large $L$, after which the $x$ (that is, the
radial) integral can be done. It is straightforward to perform these
operations using Mathematica. To zeroth order in $\epsilon$, the
results for the first and second order of WKB are given below. For
the first order WKB term, the result is
\begin{eqnarray}
&& \hspace{-2cm} \int_0^\infty dr\, \int_0^\infty ds
\left[-\frac{1}{s}\left( e^{-m^2 s}\,s^{\epsilon}\right)\right]
\left(\sum_{l=L+\frac{1}{2}}^\infty f_l^{(1)}(s,r)\right)
\nonumber \\
&=& \frac{1}{24\epsilon} +2L^2+4L- \frac{\ln
L}{2}\left(\frac{1}{6}+m^2\right) +\frac{119}{72}   - \frac{\ln
2}{12}+ \frac{\psi(\frac{1}{2})}{24}
\nonumber \\
&& + \frac{m^2}{2} \left(1-2\ln 2+\ln m\right) -\frac{1 +
6\,m^2}{12\,L} +O\left(\frac{1}{L^2}\right), \label{wkb1}
\end{eqnarray}
where $\psi(\frac{1}{2})=-\gamma-2\ln 2$. For the
second order WKB term, the result is
\begin{eqnarray}
&& \hspace{-2cm} \int_0^\infty dr\, \int_0^\infty ds
\left[-\frac{1}{s}\left( e^{-m^2 s}\,s^{\epsilon}\right)\right]
\left( \sum_{l =
L+\frac{1}{2}}^\infty f_l^{(2)}(s,r)\right) \nonumber \\
&=& \frac{1}{24\epsilon}-\frac{\ln L}{12} +\frac{1}{9}  - \frac{\ln
2}{12}+ \frac{\psi(\frac{1}{2})}{24} - \frac{1}{12\,L}
+O\left(\frac{1}{L^2}\right). \label{wkb2}
\end{eqnarray}
The third order WKB term gives a contribution of at most
$O(\frac{1}{L^2})$, and has no $\epsilon$ pole. Similarly, it can be shown that all higher order
WKB terms have no $\epsilon$ pole, and vanish for large $L$.

We can thus compute the large $L$ limit by considering only the
relevant parts of the first two WKB expressions in (\ref{wkb1}) and
(\ref{wkb2}). Inserting the identification between $\epsilon$ and
$\Lambda$ in (\ref{corr}), we obtain
\begin{eqnarray}
\sum_{l=L+\frac{1}{2}}^\infty \Gamma_{\Lambda, (l)}^S(A; m)&\sim& \frac{1}{6}\ln \Lambda+2 L^2 + 4 L-\left(\frac{1}{6}+\frac{m^2}{2}\right)\ln L \nonumber\\
&& +\left[\frac{127}{72}-\frac{1}{3}\ln 2+\frac{m^2}{2}-m^2 \ln 2+\frac{m^2}{2}\ln m \right]+O\left(\frac{1}{L}\right)
\label{div}
\end{eqnarray}
We can now identify the physical role of the various terms in
(\ref{div}). The first term is the expected logarithmic counterterm
which is subtracted in (\ref{renaction}),
and explains the origin of the $\frac{1}{6}\ln \mu$ term in (\ref{modaction}).
The next three terms give
quadratic, linear and logarithmic divergences in $L$. We shall show
in the next section that these divergences cancel corresponding
divergences in the first sum in (\ref{actionsplit}), which were
found in our numerical data. It is a highly nontrivial check on this
WKB computation that these divergent terms have the correct
coefficients to cancel these divergences.
Note that the $\ln L$ coefficient, and the finite term, are mass dependent.

Thus, the minimally subtracted renormalized effective action $\tilde{\Gamma}^S_{\rm ren}(m)$, defined in (\ref{modaction}),  is
\begin{eqnarray}
\tilde{\Gamma}^S_{\rm ren}(m)&=&\lim_{L\to\infty}\left\{\sum_{l=0,\frac{1}{2},\dots}^L \Gamma^S_l(A;m)
+2 L^2 + 4 L-\left(\frac{1}{6}+\frac{m^2}{2}\right)\ln L  \right.\nonumber \\
&&\left. +\left[\frac{127}{72}-\frac{1}{3}\ln 2+\frac{m^2}{2}-m^2 \ln 2+\frac{m^2}{2}\ln m \right]\right\}\quad ,
\label{answer}
\end{eqnarray}
where the first sum is to be computed numerically from the partial wave expansion in (\ref{lj}).

\section{Numerical Calculation}
\label{numerical}

In this section we describe the numerical technique for computing
the radial determinants which enter the partial wave expansion in
(\ref{lj}). These one-dimensional determinants can be computed
efficiently using the following result
\cite{levit,coleman,dreyfus,forman,kirsten}. Suppose ${\mathcal
M}_1$ and ${\mathcal M}_2$ are two second order ordinary
differential operators on the interval $r\in [0,\infty)$, with
Dirichlet boundary conditions assumed. Then the ratio of the
determinants of ${\mathcal M}_1$ and ${\mathcal M}_2$ is given by
\begin{eqnarray}
\det\left(\frac{{\mathcal M}_1}{{\mathcal M}_2}\right)=\lim_{R\to\infty}\left(\frac{\psi_1(R)}{\psi_2(R)}\right)
\label{theorem}
\end{eqnarray}
where $\psi_i(r)$ (for $i=1,2$) satisfies the initial value problem
\begin{eqnarray}
{\mathcal M}_i\, \psi_i(r)=0 , \qquad {\rm with}\quad \psi_i(0)=0\quad {\rm and}\quad \psi^\prime(0)=1\quad .
\label{ode}
\end{eqnarray}
Since an initial value problem is very simple to solve numerically,
this theorem provides an efficient way to compute the determinant of an ordinary differential operator.
Note in particular that no direct information about the spectrum
(either bound or continuum states, or phase shifts) is required in order to compute the determinant.

We can simplify the numerical computation further. Note that for the
free massive Klein-Gordon partial-wave operator, ${\mathcal H}^{{\rm
free}}_{(l)}+m^2$ (with ${\mathcal H}^{{\rm free}}_{(l)}$ given in
(\ref{freeh})), the solution to (\ref{ode}) is the modified Bessel
function \cite{abramowitz}
\begin{eqnarray}
\psi^{\rm free}_{(l)}(r)=\frac{I_{2l+1}(m r)}{r}\quad .
\label{besseli}
\end{eqnarray}
This solution grows exponentially fast at large $r$, as do the
numerical solutions to (\ref{ode}) for the operators ${\mathcal
H}_{(l,j)}+m^2$, with ${\mathcal H}_{(l,j)}$ specified in
(\ref{insth}). Thus, it is numerically better to consider the ODE
satisfied by the {\it ratio} of the two functions
\begin{eqnarray}
{\mathcal R}_{(l,j)}(r)=\frac{\psi_{(l,j)}(r)}{\psi^{\rm free}_{(l)}(r)}.
\label{ratio}
\end{eqnarray}
 This quantity
has a finite value in the large $r$ limit, which is just the ratio of the determinants as in (\ref{theorem}).
The boundary conditions for the ratio function are
\begin{eqnarray}
{\mathcal R}_{(l,j)}(0)=1 \qquad ; \qquad  {\mathcal R}^\prime_{(l,j)}(0)=0
\label{bcf}
\end{eqnarray}
A similar idea of considering the ratio function was used by Baacke and Lavrelashvili in their analysis of
metastable vacuum decay \cite{baacke}.

It is worthwhile making a brief side comment about the boundary conditions at $r=0$.
The two functions on the right hand side of (\ref{ratio}) do not necessarily satisfy
the boundary conditions at $r=0$ in (\ref{ode}). For example, when $l=0$, $\psi^{\rm free}_{(l=0)}(r)$ does not vanish at $r=0$. And in the massless case this issue is more serious. However, since only the ratio is important, one can introduce an ultraviolet regulator by imposing the boundary conditions in (\ref{ode}) at $r=a$, for $a$ small but nonzero. Then the free solution satisfying the boundary conditions is a linear combination of the two modified Bessel functions $\frac{I_{2l+1}(m r)}{r}$ and $\frac{K_{2l+1}(m r)}{r}$. It is straightforward to show  that the differential equation governing the ratio function,
and the asymptotic ($r\to \infty$) value of the ratio function are independent of $a$ as $a\to 0$.

In fact, since we are ultimately interested in the logarithm of the
determinant, it is more convenient (and more stable numerically) to
consider the logarithm of the ratio, i.e.,
 \begin{eqnarray}
 S_{(l,j)}(r)\equiv\ln {\mathcal R}_{(l,j)}(r)\quad ,
 \label{logfunction}
 \end{eqnarray}
which satisfies the differential equation
\begin{eqnarray}
\frac{d^2 S_{(l,j)}}{dr^2}+\left(\frac{d S_{(l,j)}}{dr}\right)^2+\left(\frac{1}{r}+2m\frac{I^\prime_{2l+1}(m r)}{I_{2l+1}(m r)}\right)\frac{d S_{(l,j)}}{dr}=U_{(l,j)}(r)\quad ,
\label{nonlinear}
\end{eqnarray}
with boundary conditions
\begin{eqnarray}
S_{(l,j)}(r=0)=0\qquad , \qquad S^\prime_{(l,j)}(r=0)=0\quad .
\label{logbc}
\end{eqnarray}
The `potential' term $U_{(l,j)}(r)$ in (\ref{nonlinear}) is given by
\begin{eqnarray}
U_{(l,j)}(r)=\frac{4(j-l)(j+l+1)}{r^2+1}-\frac{3}{(r^2+1)^2}.
\label{potential}
\end{eqnarray}

\begin{figure}[tp]
\psfrag{r}{\Large $r$}
\psfrag{l}{\Large $l$}
\psfrag{=}{\Large $\!=$}
\psfrag{0}{\Large $0$}
\psfrag{10}{\Large $10$}
\psfrag{20}{\Large $20$}
\psfrag{30}{\Large $30$}
\includegraphics[scale=1.75]{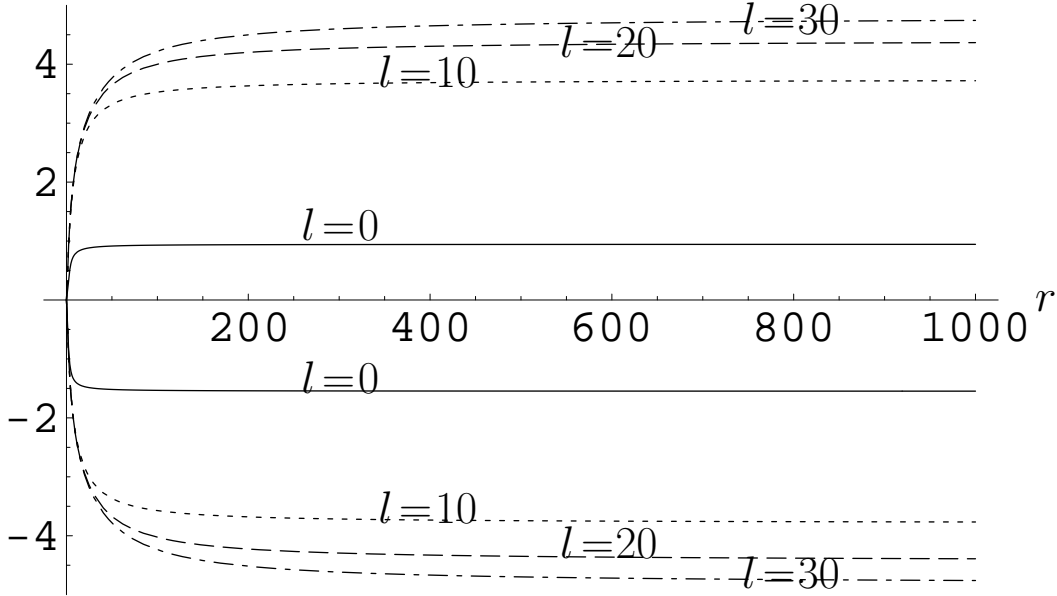}
\caption{Plots of the $r$ dependence of $S_{(l,l+\frac{1}{2})}(r)$
and $S_{(l+\frac{1}{2},l)}(r)$,
solutions of the nonlinear differential equation in (\protect{\ref{nonlinear}}),
for $m=1$, and for $l=0,10,20,30$.
The upper curves are for $S_{(l,l+\frac{1}{2})}(r)$, while the lower ones are for
$S_{(l+\frac{1}{2},l)}(r)$.
Note that the curves quickly reach an asymptotic large-$r$ 
constant value, and also notice that the contributions from
$S_{(l,l+\frac{1}{2})}(r=\infty)$ and
$S_{(l+\frac{1}{2},l)}(r=\infty)$ almost cancel one another when
summed. \label{fig2}}
\end{figure}

To illustrate the computational method, in Figure \ref{fig2} we plot
$S_{(l,l+\frac{1}{2})}(r)$ and $S_{(l+\frac{1}{2},l)}(r)$ for
various values of $l$, with mass value $m=1$ (which is in the region
in which neither the large nor small mass expansions is accurate).
Note that the curves quickly reach an asymptotic large-$r$ 
constant value, and also notice that the contributions from
$S_{(l,l+\frac{1}{2})}(r=\infty)$ and
$S_{(l+\frac{1}{2},l)}(r=\infty)$ almost cancel one another when
summed. This behavior is generic for all values of mass $m$.


To obtain very high precision for $S_{(l,j)}(r=\infty)$ in the numerical computation,
it proves useful to make a further numerical modification.
For large $r$, a good first approximation
to $S_{(l,j)}(r)$ is provided by neglecting the first two terms on the
left-hand-side of the differential equation in (\ref{nonlinear}).
Thus we define a new function $T_{(l,j)}(r)$ by
\begin{eqnarray}
S_{(l,j)}(r)&=&\int_0^r dr^\prime \left(\frac{U_{(l,j)}(r^\prime)}
{W_l(r^\prime)}\right)
+T_{(l,j)}(r), \\
W_l(r)&=&\frac{1}{r}+2m\frac{I^\prime_{2l+1}(m r)}{I_{2l+1}(m r)}.
\label{stot}
\end{eqnarray}
This  new function $T_{(l,j)}(r)$ satisfies the modified equation
\begin{eqnarray}
\frac{d^2 T_{(l,j)}}{dr^2}+\left(\frac{d T_{(l,j)}}{dr}\right)^2
+\left(W_l(r)+2\frac{U_{(l,j)}(r)}{W_l(r)}\right)\frac{d T_{(l,j)}}{dr}=
-\left(\frac{U_{(l,j)}(r)}{W_l(r)}\right)^2
-\frac{d \left(\frac{U_{(l,j)}(r)}{W_l(r)}\right)}{d r}
\label{Teq}
\end{eqnarray}
with the boundary conditions: $T_{(l,j)}(0)=T_{(l,j)}^\prime(0)=0$.
Numerical values for the quantities defined in (\ref{stot})
provide greater accuracy at large $r$, and are also better for large $l$.
In fact, we can iterate this type of transformation as many times as we wish.
For our computation we achieved excellent numerical precision by iterating
this transformation twice.


\begin{figure}[tp]
\psfrag{splus}{\Large $S_{(l,l+\frac{1}{2})}(\infty)$}
\psfrag{sminus}{\Large $S_{(l+\frac{1}{2},l)}(\infty)$}
\psfrag{total}{\Large $P(l)$}
\psfrag{hor}{\Large $l$}
\includegraphics[scale=1.7]{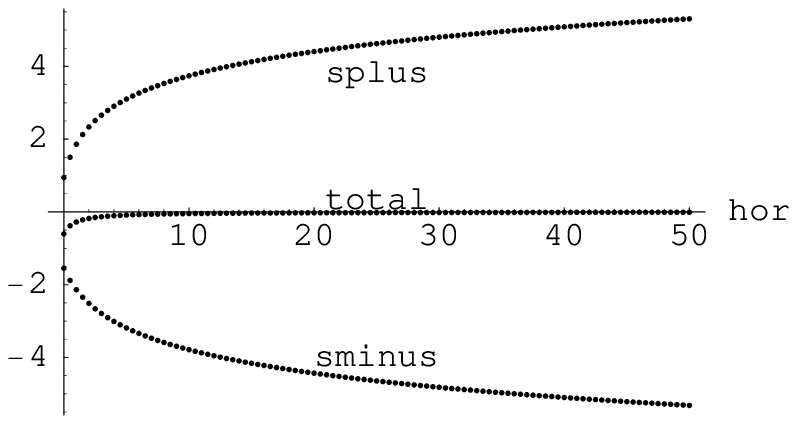}
\caption{Plot of $S_{(l,l+\frac{1}{2})}(r=\infty)$, $S_{(l+\frac{1}{2},l)}(r=\infty)$, and their sum $P(l)$, defined in (\protect{\ref{pl}}), for $m=1$. Note that $S_{(l,l+\frac{1}{2})}(r=\infty)$ and $S_{(l+\frac{1}{2},l)}(r=\infty)$ almost cancel, with their sum $P(l)$ vanishing at large $l$. See also Figure \protect{\ref{fig4}}.}
\label{fig3}
\end{figure}

\begin{figure}[tp]
\psfrag{hor}{\Large $l$}
\psfrag{ver}{\Large $P(l)$}
\includegraphics[scale=1.7]{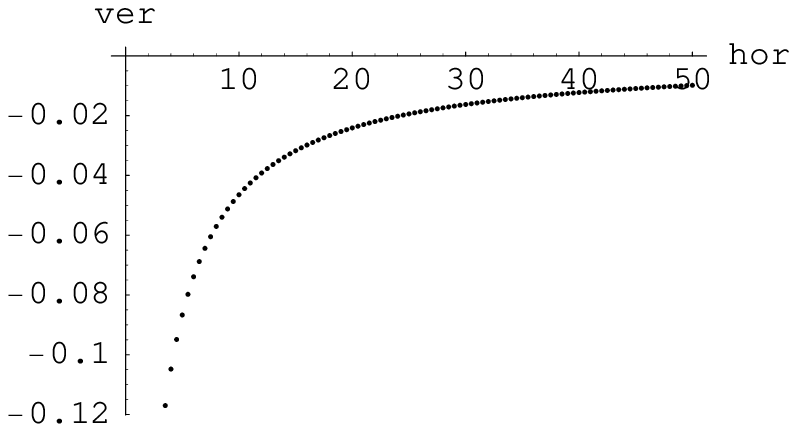}
\caption{Plot of the $l$ dependence of $P(l)$,  for $m=1$. (This is a blow-up of the $P(l)$ data from Figure \protect{\ref{fig3}}). $P(l)$ behaves like
$O(\frac{1}{l})$ for large $l$. Note that this implies that the sum over
$l$ in (\protect{\ref{lj1}}) has quadratic divergences in the large $L$ limit.
\label{fig4}}
\end{figure}

The large $r$ values of $S_{(l,l+\frac{1}{2})}(r)$ and
$S_{(l+\frac{1}{2},l)}(r)$ can be extracted with very good precision
(we integrated out to $r=10^8$). Notice that the asymptotic values
of  $S_{(l,l+\frac{1}{2})}(r)$ and $S_{(l+\frac{1}{2},l)}(r)$ very
nearly cancel one another, as illustrated in Figure \ref{fig3}. This behavior occurs for all $m$, and
becomes more accurate as $l$ increases. In fact, for a given mass,
it is found that, as a function of $l$,
$S_{(l,l+\frac{1}{2})}(r=\infty)$ grows like $\ln l$ while
$S_{(l+\frac{1}{2},l)}(r=\infty)$ decreases like $-\ln l$. This
divergence cancels in the sum, resulting in the behavior:
\begin{eqnarray}
P(l)&\equiv&S_{(l,l+\frac{1}{2})}(r=\infty)+S_{(l+\frac{1}{2},l)}(r=\infty)\nonumber\\
& \sim& O \left( \frac{1}{l} \right)\qquad , \quad l\to\infty
\label{pl}
\end{eqnarray}
This behavior is illustrated in Figure \ref{fig4}, which is a blow-up of the $P(l)$ data in Figure \ref{fig3}.

Recall from (\ref{lj}) that the first numerical sum in
(\ref{answer}) is in fact a sum over $P(l)$, with degeneracy factors:
\begin{eqnarray} \sum_{l=0,\frac{1}{2},\dots}^L \Gamma_l^S(A;
m)&=&\sum_{l=0,\frac{1}{2}, \dots}^L (2l+1)(2l+2)  P(l) \label{lj1}
\end{eqnarray}
Thus, we can rewrite our final expression
(\ref{answer}) for the minimally subtracted renormalized effective
action $\tilde{\Gamma}^S_{\rm ren}(m)$ as
\begin{eqnarray}
\tilde{\Gamma}^S_{\rm ren}(m)&=&\lim_{L\to \infty}\left\{\sum_{l=0,\frac{1}{2},\dots}^L (2l+1)(2l+2) P(l)
+2 L^2 + 4 L-\left(\frac{1}{6}+\frac{m^2}{2}\right)\ln L  \right.\nonumber \\
&&\left. +\left[\frac{127}{72}-\frac{1}{3}\ln 2+\frac{m^2}{2}-m^2
\ln 2+\frac{m^2}{2}\ln m \right]\right\}.
\label{answer2}
\end{eqnarray}
The first sum is over terms that are computed numerically, as
described above. The rest represents renormalization terms which
have been computed using minimal subtraction and WKB.

\begin{figure}[tp]
\psfrag{hor}{\Large $m$}
\psfrag{ver}{\Large $\tilde\Gamma_{\rm ren}^S(m)$}
 \includegraphics[scale=1.75]{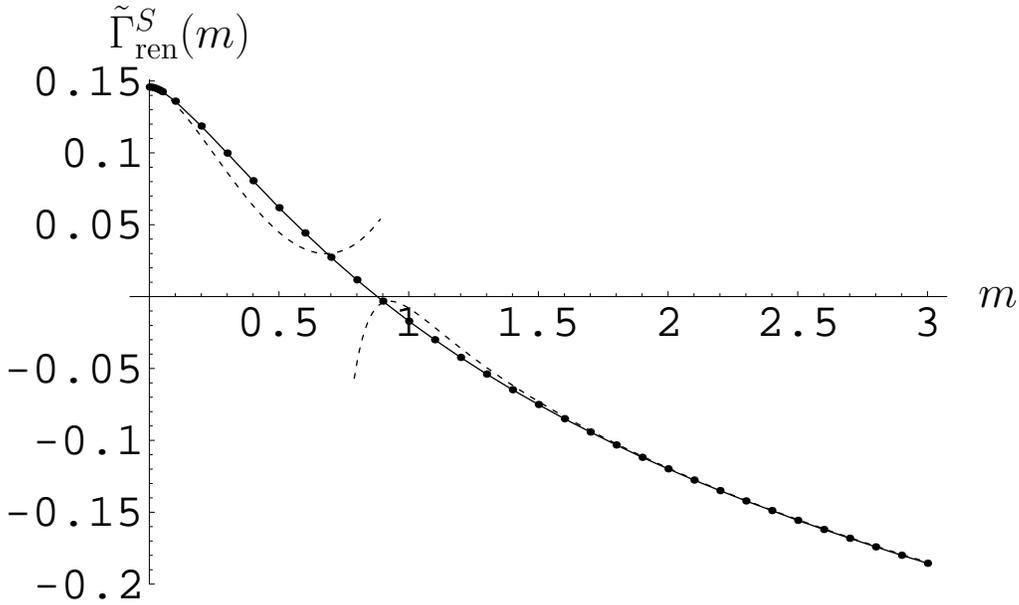}
\caption{Plot of our numerical results for $\tilde{\Gamma}^{S}_{\rm ren}(m)$
from (\protect{\ref{answer2}}), compared  with the analytic extreme small
and large mass limits
[dashed curves] from (\protect{\ref{masslimit}}).
The dots denote numerical data points from (\protect{\ref{answer2}}),
and the solid line is a fit through these points. The agreement with the analytic small and large mass limits is very precise.
\label{fig5}}
\end{figure}

Since the degeneracy factor $(2l+1)(2l+2)$ is quadratic, the large
$l$ behavior of $P(l)$ indicated in (\ref{pl}) [and plotted in Figure \ref{fig4}] shows that in the
large $L$ limit, the sum (\ref{lj1}) has potentially divergent terms
going like $L^2$, $L$ and $\ln L$, as well as terms finite and
vanishing for large $L$. Remarkably, we find that these divergent terms are
exactly canceled by the divergent large $L$ terms found in the
previous section for the second sum in (\ref{actionsplit}).
Thus, the
renormalized effective action $\tilde{\Gamma}^S_{\rm ren}(m)$
calculated using (\ref{answer2}) is finite, converges for large $L$,
and can be computed for any mass $m$.
We found excellent convergence with $L=50$ in our
numerical data, combined with Richardson extrapolation
\cite{bender}. In Figure
\ref{fig5} we plot these results for $\tilde{\Gamma}^{S}_{\rm ren}(m)$,
and compare them with the analytic
small and large mass expansions in (\ref{masslimit}). The agreement
is spectacular. Thus, our expression (\ref{answer2}) provides a
simple and numerically precise interpolation between the large mass
and small mass regimes.

As an interesting check, our formula (\ref{answer2}) provides a very
simple computation of 't Hooft's leading small mass result. In the
$m\to 0$, (\ref{nonlinear}) becomes
\begin{eqnarray}
\frac{d^2 S_{(l,j)}}{dr^2}+\left(\frac{d S_{(l,j)}}{dr}\right)^2+
\left(\frac{4l+3}{r}\right)\frac{d S_{(l,j)}}{dr}=U_{(l,j)}(r).
\label{massless}
\end{eqnarray}
One can find the solution of this equation in analytic form:
\begin{eqnarray}
S_{(l,l+\frac{1}{2})}(r)&=&\ln\left[\frac{2l+1}{2l+2} \right]+
\ln\left[\sqrt{1+r^2}+\frac{1}{2l+1}\frac{1}{\sqrt{1+r^2}}\right] \\
S_{(l+\frac{1}{2},l)}(r)&=&-\ln\left[\sqrt{1+r^2}\right].
\label{masslesssol}
\end{eqnarray}
As $r\to \infty$, each quantity diverges but
the sum, $P(l)$, has a finite value
\begin{eqnarray}
P(l)=\ln\left[\frac{2l+1}{2l+2}\right].
\label{exact}
\end{eqnarray}
Then it follows that
\begin{eqnarray}
\tilde{\Gamma}^S_{\rm ren}(m=0)&=&\lim_{L\to\infty}\left\{\sum_{l=0,\frac{1}{2},\dots}^L (2l+1)(2l+2)
\ln\left(\frac{2l+1}{2l+2}\right) +2 L^2 + 4 L-\frac{1}{6}\ln L
+\frac{127}{72}-\frac{1}{3}\ln 2\right\} \nonumber\\
&=& -\frac{17}{72}-\frac{1}{6}\ln 2 +\frac{1}{6}-2\zeta^\prime(-1)\nonumber\\
&=& \alpha\left(\frac{1}{2}\right)= 0.145873...
\label{masslessanswer}
\end{eqnarray}
which agrees precisely with the leading term (\ref{alphahalf}) in the small mass limit in (\ref{masslimit}).

\section{Comparison With Other Results}
\label{comparison}

\begin{figure}[tp]
\psfrag{hor}{\Large $m$}
\psfrag{ver}{\Large $\tilde\Gamma_{\rm ren}^S(m)$}
\includegraphics[scale=1.75]{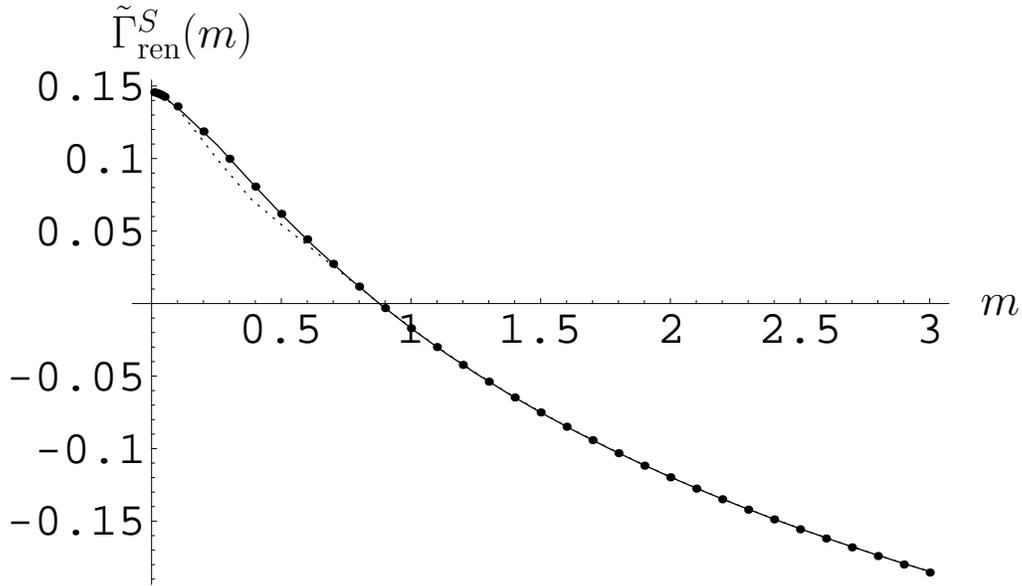}
\caption{Plot of "Pad\'e" approximations of the effective action. The dotted line is the interpolation in (\ref{interpol}) proposed in \protect{\cite{kwon}}, while the solid line is the approximation in (\ref{interpol2}) which is an interpolating function fit to the exact mass dependence found in this paper. The solid dots are the exact numerical data.}
\label{fig6}
\end{figure}

Since this is the first computation of the full mass dependence of
the one-loop effective action in an instanton background, there is
not much with which we can compare, except the small and large mass limits (\ref{masslimit}), which agree very well.
There are, however, two other comparisons worth making. The first is with a modified Pad\'e interpolating fit proposed in \cite{kwon}, which is
consistent with the two leading terms in each of the known analytic small and
large mass limits given in (\ref{masslimit}):
\begin{eqnarray}
\tilde{\Gamma}^S_{\rm ren}(m)\sim -\frac{1}{6} \ln m +\frac{\frac{1}{6}\ln m +\alpha-(3\alpha+\beta) m^2 -\frac{1}{5}
m^4}{1-3m^2+20m^4+15m^6},
\label{interpol}
\end{eqnarray}
with $\alpha\equiv\alpha(1/2)\sim0.145873$ and
$\beta=\frac{1}{2}(\ln2-\gamma)\sim 0.05797$. Based on the numerical
data found in the present paper, we can fit the exact mass
dependence in Figure \ref{fig5} with an expanded form of this
interpolating function. Let us assume the form
\begin{eqnarray}
\tilde{\Gamma}_{\rm ren}^S(m)\sim -\frac{1}{6} \ln m +\frac{\frac{1}{6}\ln m +\alpha-
(3\alpha+\beta) m^2 +A_1 m^4 -A_2 m^6}{1-3m^2+B_1 m^4+ B_2 m^6+  B_3 m^8}\quad .
\label{interpol2}
\end{eqnarray}
One may easily check that the leading two terms of the small mass expansion of this
expression (\ref{interpol2}) is the same as the small mass expansion in (\ref{masslimit}).
Then, comparing the four leading terms of the large mass expansion of (\ref{interpol2}) with  the large mass expansion of (\ref{masslimit})
fixes the coefficients $B_1, B_2, B_3$ to be:
\begin{eqnarray}
B_1&=&25\left(\frac{592955}{21609}A_2+\frac{255}{49}A_1+9\alpha+3\beta\right),  \nonumber \\
B_2&=&-75\left(\frac{85}{49}A_2+A_1\right), \quad B_3=75 A_2.
\end{eqnarray}
There remain two free parameters, $A_1$ and $A_2$, unfixed in (\ref{interpol2}). We can choose them so that
the Pad\'e approximant in (\ref{interpol2}) best fits the numerical data found in
the previous section.  This is a straightforward numerical exercise, and we find
the best fit is given by
\begin{equation}
A_1=-13.4138, \quad A_2=2.64587
\end{equation}
In Fig. \ref{fig6}, we compare these approximations with the precise
numerical data. Note that the fit based on (\ref{interpol2}) [solid
line] is extremely precise, so we can use (\ref{interpol2}) as a
simple analytic expression approximating the full mass dependence of
the effective action, over the entire range of mass values. This is analogous to modified Pad\'e fits used in chiral extrapolation of lattice data \cite{lattice}, and has also been explored for Heisenberg-Euler effective actions \cite{chiral}. 
This form will also be useful if one wishes to capture the full $\rho$-dependence (for a given quark mass value) of the scalar effective action $\tilde{\Gamma}_{\rm ren}^S(m)$ via (\ref{modaction}) (and then also of the fermion effective action $\tilde{\Gamma}_{\rm ren}^F(m)$ via (\ref{susy})). 

\begin{figure}[tp]
\psfrag{hor}{\Large $m$}
\psfrag{ver}{\Large $\tilde\Gamma_{\rm ren}^S(m)$}
 \includegraphics[scale=1.75]{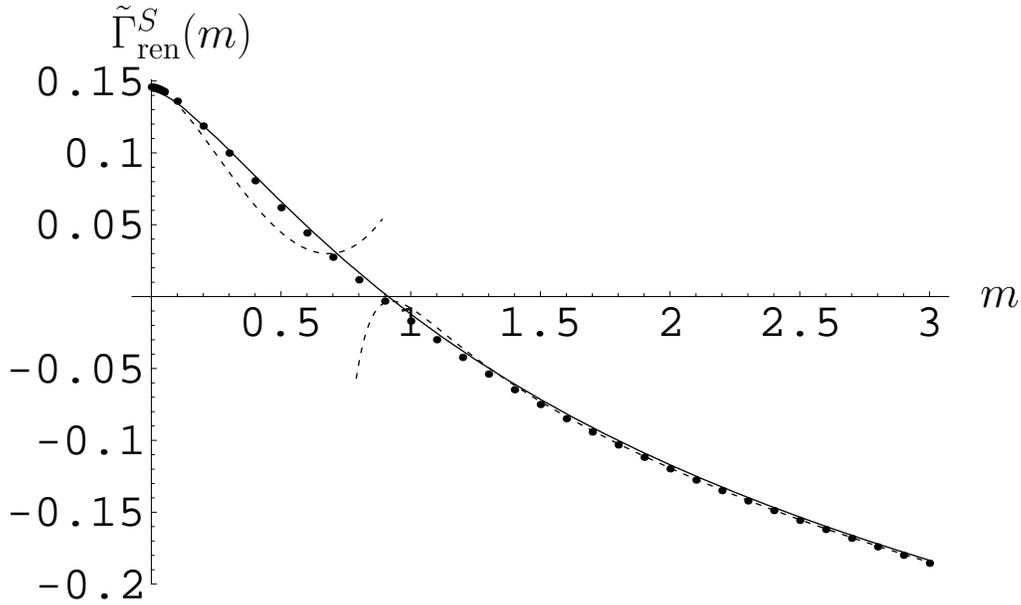}
\caption{Plot of $\tilde{\Gamma}^{S}(m)$, comparing the leading derivative expansion
approximation (solid line) with the precise numerical answers (dots). The dashed lines show the small and large mass limits from (\protect{\ref{masslimit}}).
\label{fig7}}
\end{figure}

Another comparison we can make is to the derivative expansion approximation.
This approximation was already studied in \cite{wkbpaper}, where it
was noted that it was remarkably close to the extreme small and
large mass limits in (\ref{masslimit}). Now that we have computed
the full mass dependence of $\tilde{\Gamma}^S_{\rm ren}(m)$, it is
worth revisiting this comparison. Recall that the philosophy of the
derivative expansion is to compute the one-loop effective action for
a covariantly constant background field, which can be done exactly,
and then perturb around this constant background solution. The
leading order derivative expansion for the effective action is
obtained by first taking the (exact) expression for the effective
Lagrangian in a covariantly constant background, substituting the
space-time dependent background, and then integrating over
space-time. For an instanton background, which is self-dual, we base
our derivative expansion approximation on a covariantly constant and
self-dual background \cite{dunne, ds1}. 
This leads to the following simple integral representation for the
leading derivative expansion approximation to the effective action
\cite{wkbpaper}
\begin{eqnarray}
\left.\tilde{\Gamma}^S_{\rm ren}(A; m)\right]_{\rm DE}
&=&-\frac{1}{14}\int_0^\infty \frac{dx\, x}{e^{2\pi x}-1}\left\{ -84+14 \ln\left(1+\frac{48 x^2}{m^4}\right)
+7\sqrt{3}\,\frac{m^2}{x} {\rm arctan}\left(\frac{4\sqrt{3}\, x}{m^2}\right) \right.\nn\\
&&\left. \hskip 3cm +768\frac{x^2}{m^4}~_2 F_1\left(1,\frac{7}{4},\frac{11}{4};-\frac{48x^2}{m^4}\right)\right\} -\frac{1}{6}\ln m
\label{dm}
\end{eqnarray}
Figure \ref{fig7} shows a comparison of this leading derivative
expansion expression with the exact numerical data. In the range
covered, the agreement is surprisingly good for such a crude
approximation.


\section{Concluding Remarks}
\label{conclusion}

In this paper  we have presented the details of a computation of the
fermion determinant in an instanton background for all values of the
quark mass. The agreement with the known analytic expressions in the
small and large mass limits is excellent. As another application of our result,
we can reinstate the dependence on $\rho$, the instanton scale parameter,
simply by replacing $m$ by $m \rho$. Then, given a quark field of fixed
mass $m$, the fermion determinant as a function of instanton size
$\rho$ can be studied. For phenomenological applications, this can now be
simply described with better than 1\% 
numerical accuracy by using the interpolating function
(\ref{interpol2}) for $\tilde{\Gamma}^{S}_{\rm ren}$, together with our formulas 
(\ref{modaction}) and
(\ref{susy}). The resulting interpolating expression for the fermion determinant in an instanton background is
\begin{eqnarray}
&& e^{-\Gamma^F_{\rm ren}} \label{fermiondet} \\
&& = \frac{m}{\mu} (\mu\rho)^{\frac{1}{3}} \exp \left\{ -\frac{1}{3} \ln
m\rho +\frac{\frac{1}{3}\ln (m\rho) +2\alpha- (6\alpha+2\beta)
(m\rho)^2 +2A_1 (m\rho)^4 -2A_2 (m\rho)^6}{1-3(m\rho)^2+B_1
(m\rho)^4+ B_2 (m\rho)^6+ B_3 (m\rho)^8} \right\}. \nonumber
\end{eqnarray}
This expression assumes minimal subtraction for the renormalized
coupling entering the tree-level contribution. To obtain the
corresponding expression for $e^{-\Gamma^F_{\rm ren}}$ in other
renormalization schemes, one needs to perform additional finite
renormalizations, as discussed in Refs.\cite{kwon,hasenfratz}. Notice
that such one-loop finite renormalization terms $\Delta \Gamma^F_{\rm
finite}$ can have dependence on the quark mass $m$ (but not on $\rho$),
and the lack of manifest decoupling for large quark mass
in the expression (\ref{fermiondet}) is a renormalization artifact
\cite{kwon}. We also remark that in instanton-based QCD
phenomenology one may well choose the quark mass value $m$ in
(\ref{fermiondet}) to be different from the Lagrangian (or current)
quark mass, taking instead some effective mass value
\cite{faccioli}. Our formula (\ref{fermiondet}) can be used for
discussing instanton effects in gauge theories with compact extra
dimensions as well \cite{poppitz}.

The computational method we described is versatile and can be
adapted to a large class of previously insoluble computations of
one-loop functional determinants in nontrivial backgrounds in
various dimensions of space-time, as long as the spectral problem of
the given system can be reduced to that of partial waves. One may
especially consider using analogous methods for the computation of
quantum corrections to the soliton energy in field theories. Several
examples along this direction are currently under investigation.

\vskip .5cm {\bf Acknowledgments:} GD thanks H. Gies,  V. Khemani
and K. Kirsten for helpful comments, and the US DOE for support through
the grant DE-FG02-92ER40716.
The work of JH was supported by the Brain Korea 21 Project. 
The work of CL was supported by the Korea Science Foundation ABRL
program (R14-2003-012-01002-0). HM thanks the University of
Connecticut for hospitality during sabbatical, and the University of
Seoul for support.

\newpage
\appendix
\section{One-loop Effective Action by the WKB Phase-shift Method}

WKB theory is a powerful tool for obtaining a global approximation
to the solution of a second-order ordinary differential equation \cite{bender,dunham}.
Hence one expects that it can be utilized for the approximate
calculation of a one-dimensional functional determinant
\cite{wasson, moss}. In the case of higher-dimensional functional
determinants, which are usually needed in the one-loop effective
action calculation of field theory, one can still try to use this
WKB theory if the relevant partial differential operator becomes
separable (as is often the case with rotationally-invariant
background fields). In the latter case, however, no useful result
can be derived from such analysis if one does not have an
unambiguous renormalization procedure that goes with the WKB theory.
In fact, the usual leading-order WKB theory is not sufficient
for the determinant calculation if a consistent renormalization
demands the contribution from higher-order WKB approximation to be
included. We shall see below that this is the case.

The renormalization problem mentioned above has been solved in our
earlier paper \cite{wkbpaper}, by including needed higher-order WKB
contributions within the Schwinger proper-time representation
\cite{schwinger} for the effective action. Also achieved there is a
generalization of the Schwinger-DeWitt small proper-time expansion
\cite{dewitt, lee} to the appropriate expression for arbitrary
proper-time value (in the case of rotationally-invariant background
fields only) so that one can have an approximation to the full
effective action. In this Appendix we shall briefly summarize this
development and also provide further technical details on the
formulas  stated in \cite{wkbpaper}.

In the proper-time representation (\ref{ptaction}) for the effective
action, it is the function
\begin{equation} \label{appf}
F(s) = \int d^4 x \ {\rm tr} \langle x| \left( e^{-s(-D^2
)}-e^{-s(-\partial^2 )} \right) |x \rangle,
\end{equation}
which contains the important information. Given a rotationally invariant
background field, we may utilize the phase shift analysis with
scattering solutions of the `Hamiltonian' ${\cal H} \equiv -D^2$ to
rewrite the expression (\ref{appf}). To that end we will put the
system in a large spherical box of radius $R$ (with a Dirichlet or
Neumann boundary condition at $r=R$), thus making the spectrum
discrete \cite{thooft}. In a single instanton background in
particular, we may then consider the quantum mechanical scattering
solution for each partial wave, that is,
\begin{equation} \label{appradial}
{\cal H}_{(l, j)} \psi(r) \equiv \left\{ -\frac{\partial^2}{\partial
r^2} - \frac{3}{r} \frac{\partial}{\partial r} + \frac{4l(l+1)}{r^2}
+ \frac{4(j-l)(j+l+1)}{r^2 +1} - \frac{3}{(r^2 +1)^2} \right\}
\psi(r) = k^2 \psi(r).
\end{equation}
The corresponding free Schr\"{o}dinger equation yields
\begin{equation} \label{appradialfree}
{\cal H}^{\rm free}_{(l)} \psi_0 (r) \equiv \left\{
-\frac{\partial^2}{\partial r^2} - \frac{3}{r}
\frac{\partial}{\partial r} + \frac{4l(l+1)}{r^2} \right\} \psi_0
(r) = k^2 \psi_0 (r).
\end{equation}
We are interested in the solution of (\ref{appradial}) and
(\ref{appradialfree}) that vanish as $r^{2l}$ for $r\to 0$. Then,
for large $r$, $\psi_0 (r)$ behaves as
\begin{equation}
\psi_0 (r) \sim C r^{-3/2} \cos [ k_0 (n)(r+a) ],
\end{equation}
where
\begin{equation}
k_0 (n+1) - k_0 (n) = \frac{\pi}{R} + O \left( \frac{1}{R^2}
\right),
\end{equation}
(for nonnegative integer $n$), and $a$ is a certain constant which is
not important. On the other hand, we may write the large-$r$
asymptotic behavior of the solution to (\ref{appradial}) in the form
\begin{equation}
\psi(r) \sim C r^{-3/2} \cos[ k(n) (r+a) + \eta(k(n)) ],
\end{equation}
where $\eta(k(n))$ denotes the appropriate scattering phase shift
[The related scattering matrix is give by $S(k) = e^{2i\eta(k)}$]. 
Here, because of the boundary condition at $r=R$, we may demand the
discretized momentum $k(n)$ to be related to $k_0 (n)$ above
according to
\begin{equation} \label{appkrel}
k(n)(R+a) + \eta(k(n)) = k_0 (n)(R+a).
\end{equation}
From (\ref{appkrel}) we conclude that
\begin{equation} \label{appkrel2}
k_0 (n) = k(n) + \frac{\eta(k(n))}{R} + O \left( \frac{1}{R^2}
\right).
\end{equation}

If $[k^{l ,j}(n)]^2$ and $[k^l_0 (n)]^2$ denote the energy
eigenvalues associated with the Schr\"{o}dinger equations
(\ref{appradial}) and (\ref{appradialfree}), respectively, the
function $F(s)$ give by (\ref{appf}) can be represented as
\begin{equation} \label{appf2}
F(s) = \sum_{l=0,\frac{1}{2},\cdots} \sum_j (2l+1)(2j+1) \sum_n
\left\{ e^{-s [k^{l, j}(n)]^2} - e^{-s [k^l_0 (n)]^2} \right\},
\end{equation}
including the degeneracy factor $(2l+1)(2j+1)$ for the $(l, j)$
partial wave. But, because of (\ref{appkrel2}), we find for large
$R$
\begin{equation} \label{appkrel3}
\left\{ e^{-s [k^{l, j}(n)]^2} - e^{-s [k^l_0 (n)]^2} \right\} =
e^{-s [k^{l, j}(n)]^2} \left\{ \frac{2k^{l, j}(n)\eta_{l,
j}(k(n))}{R} s + O \left( \frac{1}{R^2} \right) \right\}.
\end{equation}
Using (\ref{appkrel3}) in (\ref{appf2}) gives rise to
\begin{equation}
F(s) \sim \sum_{l=0,\frac{1}{2},\cdots} \sum_j (2l+1)(2j+1) \sum_n
2(\Delta k) e^{-s [k^{l, j}(n)]^2} \frac{2k^{l, j}(n)\eta_{l,
j}(k(n))}{\pi} s,
\end{equation}
where $\Delta k \equiv \frac{\pi}{R}$, and then, replacing the sum
$\sum_n$ by an integral for $R\to\infty$, we obtain the following
formula:
\begin{equation} \label{appfinalf}
F(s) = \frac{2s}{\pi} \sum_{l=0,\frac{1}{2},\cdots} \sum_j
(2l+1)(2j+1) \int_0^\infty dk \ e^{-k^2 s} k \ \eta_{l, j}(k).
\end{equation}

With (\ref{appfinalf}) some caution must be exercised in dealing
with the infinite partial wave sum. Actually, in the instanton
background we are considering, the nature of the scattering problem
as defined by (\ref{appradial}) and (\ref{appradialfree}) does not
allow us to consider the $l$-sum and $j$-sum in (\ref{appfinalf}) in
a completely independent manner. The point is that, as one looks at
the given forms of ${\cal H}_{(l, j)}$ and ${\cal H}^{\rm free}_{(l)}$,
their small-$r$ behaviors match for a given $l$-value; but it is the
$j$-value that governs the large-$r$ behavior of the potential
entering ${\cal H}_{(l, j)}$, while $j$ does not appear in ${\cal
H}^{\rm free}_{(l)}$ at all. To obtain a convergent expression from
(\ref{appfinalf}), it is necessary \cite{wkbpaper} to consider
the $(l, j=l+\frac{1}{2})$ and $(l+\frac{1}{2}, j=l)$ partial-wave
contributions, both of which have the same degeneracy factor of
$(2l+1)(2l+2)$, together as one package. With this understanding, the
expression (\ref{appfinalf}) can now be cast in the form
\begin{equation} \label{appjsummedf}
F(s) = \frac{2s}{\pi} \sum_{l=0,\frac{1}{2},\cdots} (2l+1)(2l+2)
\int_0^\infty dk \ e^{-k^2 s} k \left[ \eta_{l, l+\frac{1}{2}}(k) +
\eta_{l+\frac{1}{2}, l}(k) \right].
\label{fsum}
\end{equation}
If one has complete phase shifts for all partial waves at hand, one
may use this formula (\ref{fsum}) to calculate the function $F(s)$ and then the
one-loop effective action as well. But, in the massive case, the exact phase
shifts cannot be obtained analytically. Therefore, in
\cite{wkbpaper}, we proposed to use the WKB expressions for the
phase shifts, together with our formula (\ref{appjsummedf}). This method is
elaborated below.

First, we need the results of Dunham \cite{dunham} for higher-order
WKB approximations to the scattering phase shifts. If the
Schr\"{o}dinger equation is written in the form
\begin{equation} \label{appdunhameq}
\left\{ \frac{d^2}{dx^2}+ Q(x) \right\} \Psi(x) = 0,
\end{equation}
the phase shift in the leading WKB approximation is given by
\begin{equation} \label{appeta1}
\eta^{(1)} = \frac{1}{2} \left[ \oint \sqrt{Q(x)} \ dx -
(\text{`free'}) \right],
\end{equation}
where the integration path goes around the turning point $r_1$
(i.e., the point where $Q(x)$ vanishes) in the complex plane, and
crosses the real axis at $r=r_0$ (with $r_0 <r$) and $r=r_2$ (with
$r_2$ taken to positive infinity). The choice of $r_0$ has no effect
on the value of the integral, and (`free') in (\ref{appeta1})
represents the same integral but with $Q(x)$ of the free
Schr\"{o}dinger equation. Dunham also derived the formulas in the
second and third order WKB approximations:
\begin{eqnarray}
\eta^{(2)} &=& -\frac{1}{2} \left[ \oint \frac{1}{48}
\frac{Q''(x)}{Q(x)^{3/2}} \ dx - (\text{`free'}) \right], \\
\eta^{(3)} &=& \frac{1}{2} \left[ \oint \left( \frac{1}{768}
\frac{Q^{(4)}(x)}{Q(x)^{5/2}} - \frac{7}{1536}
\frac{[Q''(x)]^2}{Q(x)^{7/2}} \ dx \right) - (\text{`free'}) \right].
\end{eqnarray}

One cannot use Dunham's formula directly with the \emph{radial}
Schr\"{o}dinger equation in (\ref{appradial}) and
(\ref{appradialfree}). The latter should be transformed
appropriately, following Langer \cite{langer}. Thus, writing $r=e^x$
and introducing the function
\begin{equation}
\Psi(x) = \left. r \psi(r) \right|_{r=e^x} =
e^{ x} \psi(r=e^x),
\end{equation}
we recast (\ref{appradial}) as
\begin{equation}
\frac{d^2 \Psi(x)}{dx^2} + e^{2x} \left\{ k^2 - \frac{4 \left( l +
\frac{1}{2} \right)^2}{e^{2x}} - \frac{4(j-l)(j+l+1)}{e^{2x} +1} +
\frac{3}{(e^{2x} +1)^2} \right\} \Psi(x) = 0.
\end{equation}
Since this is of the form (\ref{appdunhameq}), we can get the
relevant scattering phase shifts simply by setting
\begin{equation}
Q_{(l, j)}(x) = e^{2x} \left\{ k^2 - \frac{4 \left( l + \frac{1}{2}
\right)^2}{e^{2x}} - \frac{4(j-l)(j+l+1)}{e^{2x} +1} +
\frac{3}{(e^{2x} +1)^2} \right\},
\end{equation}
and also, in connection with the free equation (\ref{appradialfree}),
\begin{equation}
Q_l (x) = e^{2x} \left\{ k^2 - \frac{4 \left( l + \frac{1}{2} 
\right)^2}{e^{2x}} \right\}.
\end{equation}
Then, using the original variable $r=e^x$ and integrating by parts,
the desired WKB expressions for the phase shifts assume the form
\begin{eqnarray}
\eta_{l, j}^{(1)} &=& \frac{1}{2} \oint \sqrt{ k^2 - \tilde{V}_{(l,
j)}(r)} \ dr - (\text{`free'}), \label{appreta1} \\
\eta_{l, j}^{(2)} &=& \frac{1}{2} \oint \left\{ \frac{1}{8r^2}
\frac{1}{(k^2 - \tilde{V}_{(l, j)}(r))^{1/2}} + \frac{1}{48}
\frac{d^2 \tilde{V}_{(l, j)}(r)}{dr^2} \frac{1}{(k^2 -
\tilde{V}_{(l, j)}(r))^{3/2}} \right\} dr - (\text{`free'}),
\label{appreta2} \\
\eta_{l, j}^{(3)} &=& \frac{1}{2}\oint \left\{ -\frac{5}{128r^4}
\frac{1}{(k^2 - \tilde{V}_{(l, j)}(r))^{3/2}} - \frac{1}{128r^2}
\frac{d^2 \tilde{V}_{(l, j)}(r)}{dr^2} \frac{1}{(k^2 -
\tilde{V}_{(l, j)}(r))^{5/2}} \right. \nonumber \\
&& \left. -\frac{7}{1536} \left( \frac{d^2 \tilde{V}_{(l,
j)}(r)}{dr^2} \right)^2 \frac{1}{(k^2 - \tilde{V}_{(l,
j)}(r))^{7/2}} - \frac{1}{768} \frac{d^4 \tilde{V}_{(l,
j)}(r)}{dr^4} \frac{1}{(k^2 - \tilde{V}_{(l, j)}(r))^{5/2}} \right\}
dr \nonumber \\
&& - (\text{`free'}), \label{appreta3}
\end{eqnarray}
where $\tilde{V}_{(l, j)}(r)$ is the so-called Langer modification \cite{langer}
of the potential
\begin{equation}
\tilde{V}_{(l, j)}(r) \equiv \frac{4 \left( l + \frac{1}{2}
\right)^2}{r^2} + \frac{4(j-l)(j+l+1)}{r^2 +1} - \frac{3}{(r^2
+1)^2},
\end{equation}
and the term referred to (`free') denotes the integral expression
appearing before but with $\tilde{V}_{(l, j)}(r)$ replaced by
\begin{equation}
\tilde{V}_l (r) \equiv \frac{4 \left( l + \frac{1}{2} 
\right)^2}{r^2}.
\end{equation}
Note that $Q_{(l, j)}(x) = r^2 \left. \left\{ k^2 - \tilde{V}_{(l,
j)}(r) \right\} \right|_{r=e^x}$, and $Q^0_l (x) = r^2 \left.
\left\{ k^2 - \tilde{V}_l (r) \right\} \right|_{r=e^x}$. 

Using the results (\ref{appreta1})-(\ref{appreta3}) in
(\ref{appjsummedf}), we can simplify the expression by carrying out
the $k$-integration. First, consider the leading order WKB. The
first contour integral in (\ref{appreta1}) can be changed to the
integral along the real axis over the interval $(r_1 (k), \infty)$,
where $r_1 (k)$ is a turning point, i.e., $\left[ k^2 -
\tilde{V}_{(l, j)}(r_1 (k)) \right] = 0$. On the other hand,
$k$-integral in (\ref{appjsummedf}) runs from 0 to $\infty$. We may
here change the order of integration, that is, perform the
$k$-integral prior to considering the $r$-integration: in this case,
the integration range for $k$ would be over the interval $(k_1 (r),
\infty)$, where $k_1 (r)$ represents the value specified by the
condition $\left[ k_1 (r)^2 - \tilde{V}_{(l, j)}(r) \right] = 0$ for
a given value of $r$. Similar consideration may be given to the
second contour integral in (\ref{appreta2}), the `free' part. Then,
observing that we obtain from the first contour integral
\begin{eqnarray}
\int_{\sqrt{\tilde{V}_{(l, j)}}}^\infty dk \ \frac{2s}{\pi} \
e^{-k^2 s} \ k \ \sqrt{k^2 - \tilde{V}_{(l, j)}(r)} &=& \frac{1}{2}
\oint dk \ \frac{2s}{\pi} \ e^{-k^2 s} \ k \ \sqrt{k^2 -
\tilde{V}_{(l, j)}(r)}
\nonumber \\
&=& \frac{e^{-s \tilde{V}_{(l, j)}(r)}}{2 \sqrt{\pi} \sqrt{s}},
\end{eqnarray}
(and similarly the form $\frac{e^{-s \tilde{V}_{l}(r)}}{2 \sqrt{\pi} 
\sqrt{s}}$ from the second contour integral), we are led to the
following leading-order WKB expression for $F(s)$:
\begin{eqnarray}
&& F^{(1)}(s) = \sum_{l=0,\frac{1}{2},\cdots} (2l+1)(2l+2)
\int_0^\infty dr \left[ f^{(1)}_{(l, l+\frac{1}{2})}(s, r) +
f^{(1)}_{(l+\frac{1}{2}, l)}(s, r) \right], \label{appsmallf} \\
&& f^{(1)}_{(l, j)}(s, r) = \frac{e^{-s \tilde{V}_{(l, j)}(r)}}{2
\sqrt{\pi} \sqrt{s}} - \frac{e^{-s \tilde{V}_l (r)}}{2 \sqrt{\pi} 
\sqrt{s}}.
\end{eqnarray}
We use similar procedures to simplify the contributions coming from
the second- and third- order WKB phase shifts in (\ref{appreta2})
and (\ref{appreta3}). For this, a particularly useful relation is
\begin{equation}
\frac{1}{2} \oint dk \ \frac{2s}{\pi} \ e^{-k^2 s} \frac{k}{ \left[
k^2 - \tilde{V}(r) \right]^{n+\frac{1}{2}}} = \frac{e^{-s
\tilde{V}(r)} s^{n+\frac{1}{2}} \Gamma(-n + \frac{1}{2})}{\pi}, \
(n=0,1,2,\cdots).
\end{equation}
As a result, we obtain the higher order WKB expressions for $F(s)$,
i.e., $F^{(2)}(s)$ and $F^{(3)}(s)$, which may be expressed again by
the form (\ref{appsmallf}) but with
\begin{eqnarray}
f^{(2)}_{(l, j)}(s, r) &=& \frac{e^{-s \tilde{V}_{(l, j)}(r)}}{2
\sqrt{\pi} \sqrt{s}} \left\{ \frac{s}{4r^2} - \frac{s^2}{12}
\frac{d^2 \tilde{V}_{(l, j)}(r)}{dr^2} \right\} -\frac{e^{-s
\tilde{V}_l (r)}}{2 \sqrt{\pi} \sqrt{s}} \left\{ \frac{s}{4r^2} -
\frac{s^2}{12} \frac{d^2 \tilde{V}_l (r)}{dr^2} \right\} \\
f^{(3)}_{(l, j)}(s, r) &=&  \frac{e^{-s \tilde{V}_{(l, j)}(r)}}{2
\sqrt{\pi} \sqrt{s}} \left\{ \frac{5s^2}{32r^4} - \frac{s^3}{48r^2}
\frac{d^2 \tilde{V}_{(l, j)}(r)}{dr^2} + \frac{7s^4}{1440} \left(
\frac{d^2 \tilde{V}_{(l, j)}(r)}{dr^2} \right)^2 - \frac{s^3}{288}
\frac{d^4 \tilde{V}_{(l, j)}(r)}{dr^4} \right\} \nonumber \\
&& \hspace{-1cm} -\frac{e^{-s \tilde{V}_l (r)}}{2 \sqrt{\pi}
\sqrt{s}} \left\{ \frac{5s^2}{32r^4} - \frac{s^3}{48r^2} \frac{d^2
\tilde{V}_l (r)}{dr^2} + \frac{7s^4}{1440} \left( \frac{d^2
\tilde{V}_l (r)}{dr^2} \right)^2 - \frac{s^3}{288} \frac{d^4
\tilde{V}_l (r)}{dr^4} \right\}.
\end{eqnarray} 

In connection with using the formula (\ref{appsmallf}), our
discussion will not be complete without being clear about the
\emph{order} between executing the infinite series sum over $l$ and
performing the (improper) radial integral. This is a subtle point,
and one possible way to settle the issue unambiguously would be to
check explicitly which order  gives rise to the known small-$s$
behavior for the function $F(s)$ correctly. As was asserted in
\cite{wkbpaper}, doing the $l$-sum before the $r$-integration yields
the correct result. If instead one performs the $r$-integration first and
then considers the $l$-sum, it gives a result differing from the
correct small $s$ expression of $F(s)$ by $\frac{1}{4s}$. [Note that, although
we used the WKB series for the calculation, the thus-found
difference is an exact result since the order-dependent ambiguity is
purely a high-energy phenomenon and the WKB series can be trusted in
the high energy limit]. In view of this remark, the correct formula
to be used in the WKB analysis of the effective action should read
\begin{equation}
F(s) = \int_0^\infty dr \sum_{l=0,\frac{1}{2},\cdots} (2l+1)(2l+2)
\left[ f_{(l, l+\frac{1}{2})}(s, r) + f_{(l+\frac{1}{2}, l)}(s, r)
\right].
\end{equation}
The result of using this formula for the fermion determinant
in a single instanton background is presented in  \cite{wkbpaper}.

\section{The instanton determinant in the singular gauge}
\label{sing}

In this Appendix we address the question of the gauge invariance of the determinant or the effective action. The proper time representation of the effective action in (\ref{ptaction}) is written in terms of covariant derivatives and is clearly gauge invariant. However, to formulate the partial wave expansion, as discussed in Section \ref{radial}, we chose a particular gauge (\ref{instanton}). Thus the partial wave expansion used in the main text does not possess manifest gauge invariance. In this Appendix we verify gauge independence by showing that we obtain precisely the same result in a different gauge. Since our computational method relies on the radial symmetry of the background field, we are restricted in which gauge we can choose. The choice in (\ref{instanton}) is often called the ``regular'' gauge. But an instanton background also has radial symmetry in the so-called ``singular'' gauge:
\begin{eqnarray}
A^{\rm sing}_{\mu}(x) \equiv A_{\mu}^{a}(x)\frac{\tau^{a}}{2}= \frac{\bar\eta_{\mu\nu
a}\tau^{a}x_{\nu}\rho^2}{r^2(r^2+\rho^2)}
\label{singular}
\end{eqnarray}
in the singular gauge. In (\ref{singular}), $\bar\eta_{\mu\nu a}$
differs from $\eta_{\mu\nu a}$ in (\ref{instanton}) only by the sign in the components with $\mu$ or $\nu$ equal to 4 \cite{thooft,schaefer,shifman}.
The gauge field in the singular gauge has singular behavior in the vicinity of $r=0$.
One may worry about the validity of  the radial approach in the main text
in the singular gauge because of this.
However, it turns out that it is not so singular. Physically, the reason for this is the conformal invariance of the instanton background \cite{thooft,shifman,schaefer}.
The differential operator $-D^2$ in the instanton background (\ref{singular})
in the singular gauge can be written as, setting $\rho=1$,
\begin{equation}
-D^2_{\rm sing} \equiv \left[ - \frac{\partial^2}{\partial
r^2}-\frac{3}{r}\frac{\partial}{\partial
r}+\frac{4L^2}{r^2}+\frac{4(J^2-L^2-T^2)}{r^2(r^2+1)}+\frac{4T^2}{r^2(r^2+1)^2}
\right]\, ,
\label{singh} \end{equation}
In the region of $r\sim0$,
the "potential" term in (\ref{singh}) has $1/r^2$ singular behavior.
But the last two singular terms proportional to $T^2$ combine into
$T^2/(r^2+1)^2$, which is regular.
We decompose $1/(r^2(r^2+1))$ into $1/r^2 -1/(r^2+1)$; i.e., into singular  and regular parts.
The singular part, $(J^2-L^2)/r^2$ combines with the orbital part $L^2/r^2$.
We can cast (\ref{singh}) in the form
\begin{equation}
-D^2_{\rm sing} \equiv \left[ - \frac{\partial^2}{\partial
r^2}-\frac{3}{r}\frac{\partial}{\partial
r}+\frac{4J^2}{r^2}+\frac{4(L^2-J^2)}{(r^2+1)}-\frac{4T^2}{(r^2+1)^2}
\right]\, .
\label{singh2} \end{equation}
This should be compared with the corresponding operator in the regular gauge [see (\ref{insth})]:
\begin{equation}
-D^2_{\rm reg} \equiv \left[ - \frac{\partial^2}{\partial
r^2}-\frac{3}{r}\frac{\partial}{\partial
r}+\frac{4L^2}{r^2}+\frac{4(J^2-L^2)}{(r^2+1)}-\frac{4T^2}{(r^2+1)^2}
\right]\, .
\label{regular}\end{equation}
Note that these two have the same form except that $J^2$ and $L^2$ are
interchanged. Therefore. in the partial wave analysis
the radial Hamiltonian (for isospin $\frac{1}{2}$) can be written as
\begin{equation}
 {\cal H}^{\rm sing}_{(l,j)} \equiv \left[ - \frac{\partial^2}{\partial
r^2}-\frac{3}{r}\frac{\partial}{\partial
r}+\frac{4j(j+1)}{r^2}+\frac{4(l-j)(j+l+1)}{r^2+1}-\frac{3}{(r^2+1)^2}
\right]\, .
\label{radhsing}
\end{equation}
There is a one-to-one correspondence between the eigenvalues for the sector $(j,l)$ in the singular gauge  and those for the sector $(l,j)$ in the regular gauge, and they have the same multiplicities,  $(2j+1)(2l+1)$.

The regularized one-loop effective action in the singular gauge can be written as
\begin{eqnarray}
\Gamma_\Lambda^S(A_{\rm sing}; m)  &=& \sum_{l=0,\frac{1}{2}, \dots} (2l+1)(2l+2) \left\{ \ln \det \left(\frac{{\mathcal H}^{\rm sing}_{(l,l+\frac{1}{2})}+m^2}{{\mathcal H}^{{\rm free}}_{(l+\frac{1}{2})}+m^2}\right)+
  \ln \det \left(\frac{{\mathcal H}^{\rm sing}_{(l+\frac{1}{2},l)}+m^2}{{\mathcal H}^{{\rm free}}_{(l)}+m^2}\right)\right.
\nonumber\\
&&\hskip 2cm \left. - \ln \det \left(\frac{{\mathcal H}^{\rm sing}_{(l,l+\frac{1}{2})}+\Lambda^2}{{\mathcal H}^{{\rm free}}_{(l+\frac{1}{2})}+\Lambda^2}\right)-  \ln \det \left(\frac{{\mathcal H}^{\rm sing}_{(l+\frac{1}{2},l)}
+\Lambda^2}{{\mathcal H}^{{\rm free}}_{(l)}+\Lambda^2}\right)\right\} .
\label{pvsingular}
\end{eqnarray}
Here we have combined the radial determinants for $(l,j=l+\frac{1}{2})$ and
$(l+\frac{1}{2}, j=(l+\frac{1}{2})-\frac{1}{2})$, as in Section \ref{radial}, and we have arranged the free determinants appropriately. But from the above arguments, we know that
\begin{equation}
{\mathcal H}^{\rm sing}_{(l,l^\prime)}
= {\mathcal H}^{\rm reg}_{(l^\prime,l)}.
\end{equation}
Hence the singular gauge expression in (\ref{pvsingular}) is identical with the regular gauge expression  in (\ref{pv}).
So the Pauli-Villars regularized one loop effective action has the same
value in the singular gauge and in the regular gauge, and therefore the renormalized effective
action (\ref{renaction}), and the modified effective action (\ref{modaction}),
has the same value in each gauge.

\end{document}